\documentclass[a4paper,11pt]{article}
\usepackage{pos}

\title{Studies of the ambient light of deep Baikal waters with Baikal-GVD}
\ShortTitle{Baikal water luminescence}

\author[a]{V.M.~Aynutdinov}
\author[b]{V.A.~Allakhverdyan}
\author[a]{A.D.~Avrorin}
\author[a]{A.V.~Avrorin}
\author[c,d]{Z.~Barda\v{c}ov\'{a}}
\author[b]{I.A.~Belolaptikov}
\author[a]{E.A.~Bondarev}
\author[b]{I.V.~Borina}
\author[e]{N.M.~Budnev}
\author[l]{V.A.~Chadymov}
\author[f]{A.S.~Chepurnov}
\author[b,g]{V.Y.~Dik}
\author[a]{G.V.~Domogatsky}
\author[a]{A.A.~Doroshenko}
\author*[c]{R.~Dvornick\'{y}}
\author[e]{A.N.~Dyachok}
\author[a]{Zh.-A.M.~Dzhilkibaev}
\author[c,d]{E.~Eckerov\'{a}}
\author[b]{T.V.~Elzhov}
\author[d]{L.~Fajt}
\author[l]{V.N. Fomin}
\author[e]{A.R.~Gafarov}
\author[a]{K.V.~Golubkov}
\author[b]{N.S.~Gorshkov}
\author[e]{T.I.~Gress}
\author[h]{K.G.~Kebkal}
\author[a]{I.V.~Kharuk}
\author[b]{E.V.~Khramov}
\author[b]{M.M.~Kolbin}
\author[i]{S.O.~Koligaev}
\author[b]{K.V.~Konischev}
\author[b]{A.V.~Korobchenko}
\author[a]{A.P.~Koshechkin}
\author[f]{V.A.~Kozhin}
\author[b]{M.V.~Kruglov}
\author[j]{V.F.~Kulepov}
\author[e]{Y.E.~Lemeshev}
\author[a,\dagger]{M.B.~Milenin}
\author[e]{R.R.~Mirgazov}
\author[b]{D.V.~Naumov}
\author[f]{A.S.~Nikolaev}
\author[a]{D.P.~Petukhov}
\author[b]{E.N.~Pliskovsky}
\author[k]{M.I.~Rozanov}
\author[e]{E.V.~Ryabov}
\author[a]{G.B.~Safronov}
\author[b,g]{D.~Seitova}
\author[b]{B.A.~Shaybonov}
\author[a]{M.D.~Shelepov}
\author[a]{S.D.~Shilkin}
\author[f]{E.V.~Shirokov}
\author[c,d]{F.~\v{S}imkovic}
\author[b]{A.E.~Sirenko}
\author[f]{A.V.~Skurikhin}
\author[b]{A.G.~Solovjev}
\author[b]{M.N.~Sorokovikov}
\author[d]{I.~\v{S}tekl}
\author[a]{A.P.~Stromakov}
\author[a]{O.V.~Suvorova}
\author[e]{V.A.~Tabolenko}
\author[b]{B.B.~Ulzutuev}
\author[b]{Y.V.~Yablokova}
\author[a]{D.N.~Zaborov}
\author[b]{S.I.~Zavyalov}
\author[b]{D.Y.~Zvezdov}

\affiliation[a]{Institute for Nuclear Research, Russian Academy of Sciences, Moscow, 117312, Russia}
\affiliation[b]{Joint Institute for Nuclear Research, Dubna, 141980, Russia}
\affiliation[c]{Comenius University, 81499 Bratislava, Slovakia}
\affiliation[d]{Czech Technical University in Prague, Institute of Experimental and Applied Physics, 11000 Prague, Czech Republic}
\affiliation[e]{Irkutsk State University, Irkutsk, 664003, Russia}
\affiliation[f]{Skobeltsyn Institute of Nuclear Physics, Moscow State University, Moscow, 119991, Russia}
\affiliation[g]{Institute of Nuclear Physics of the Ministry of Energy of the Republic of Kazakhstan, Almaty, 050032, Kazakhstan}
\affiliation[h]{LATENA, St. Petersburg, 199106, Russia}
\affiliation[i]{INFRAD, Dubna, 141981, Russia}
\affiliation[j]{Nizhny Novgorod State Technical University, Nizhny Novgorod, 603950, Russia}
\affiliation[k]{St.~Petersburg State Marine Technical University, St.~Petersburg, 190008, Russia}
\affiliation[l]{Moscow, free researcher}

\note[\textdagger]{Deceased.}

\emailAdd{dvornicky@fmph.uniba.sk}

\abstract{The Baikal-GVD neutrino detector is a deep-underwater neutrino telescope under construction and recently after the winter 2023 deployment it consists of 3\,456 optical modules attached on 96 vertical strings. This 3-dimensional array of photo-sensors allows to observe ambient light in the vicinity of the Baikal-GVD telescope that is associated mostly with water luminescence. Results on time and space variations of the luminescent activity are reviewed based on data collected in 2018-2022.}

\ConferenceLogo{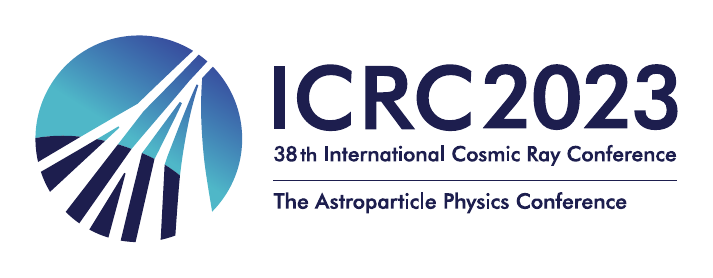}

\FullConference{%
38th International Cosmic Ray Conference (ICRC2023)\\
  26 July - 3 August, 2023\\
  Nagoya, Japan}

\begin{document}
\maketitle

\section{Introduction}

Southern part of Lake Baikal is the site for the Baikal-GVD neutrino telescope, a facility located about 3.6\,km offshore at 51$^\circ$ 46'N and 104$^\circ$ 24'E. In this region, the lakebed reaches its plateu of a nearly constant depth of 1\,366 m. The main aim of the experiment is to detect the Cherenkov radiation emitted by secondary charged particles, created in the reactions of neutrinos with surrounding medium, that are passing through the deep water in Lake Baikal. The ultimate goal is to reveal the astrophysical sources of the most energetic particles in the Universe which remain unknown yet. 

The ambient light is registered, along with the Cherenkov radiation, by photo-sensitive detectors called optical modules. The amount of the registered ambient light is derived from the noise rates recorded by the photo-multiplier tubes placed in each individual optical module. It is worth to note that the origin of the background noise rates is associated mainly with the luminescence of organic substances present in Baikal water.

In the track and cascade reconstruction analyses the ambient light plays important role as it is an unavoidable background to registered Cherenkov light. In this contribution, some selected results on water luminescence in Lake Baikal observed with the Baikal-GVD neutrino telescope are presented.
 
\section{Baikal-GVD detector design}

The basic detection unit of the Baikal-GVD telescope is optical module (OM), which comprises a 10-inch high-quantum-efficiency photo-multiplier tube (PMT) and some additional electronics equipment. 
\begin{figure}[h!]
\begin{center}$
\begin{array}{cc}
\includegraphics[width=75mm]{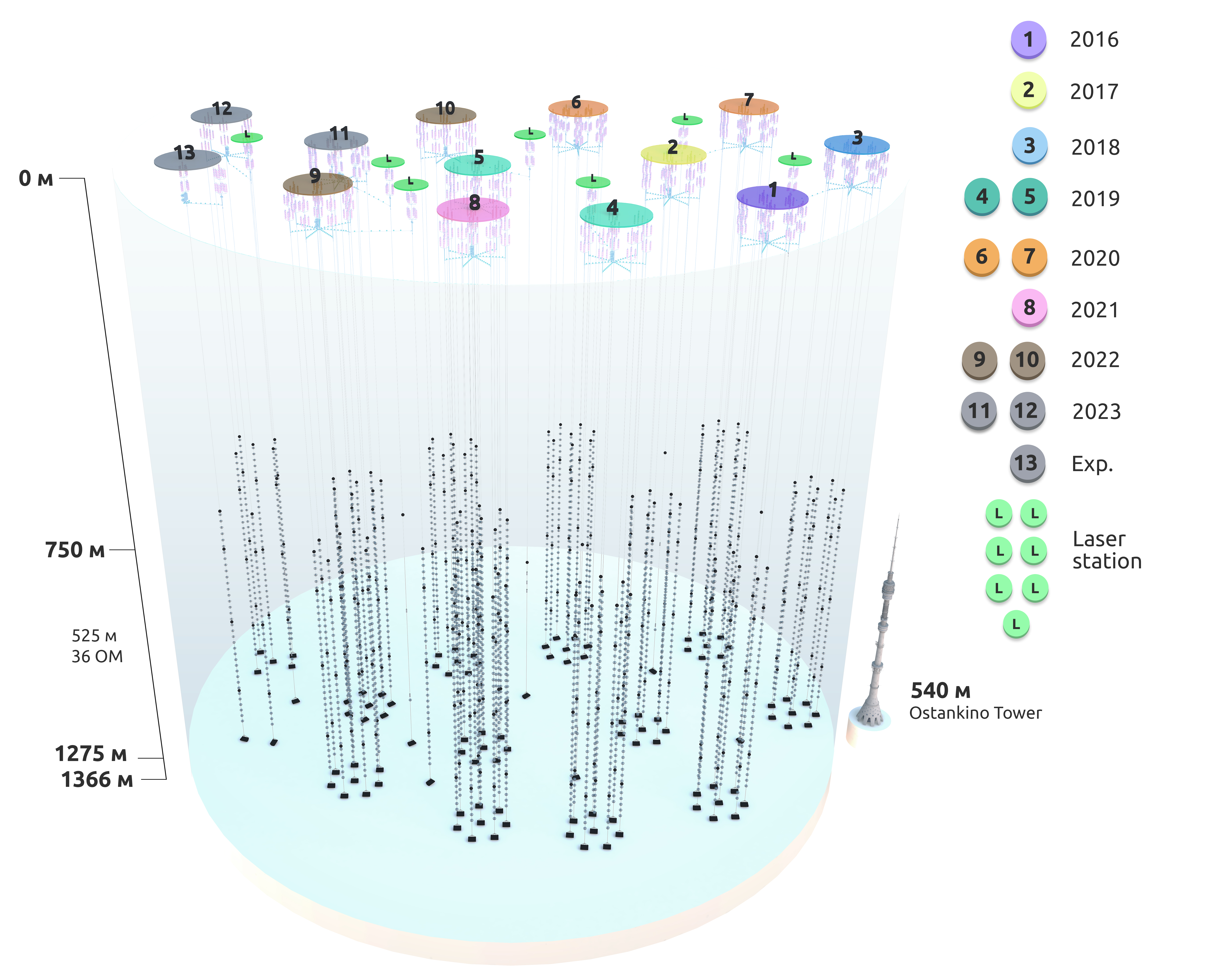}  &
\includegraphics[width=50mm]{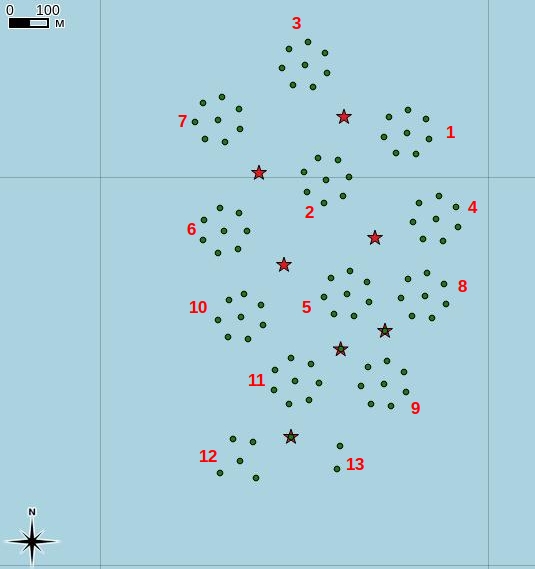}
\end{array}$
\end{center}
\caption{Left panel: Illustration of the Baikal-GVD detector after the winter deployment in the year 2023. The legend shows annual progress in the deployment. Right panel: Schematic view from the top of the Baikal-GVD detector in 2023.  
\label{fig.GVDscheme}}
\end{figure}
There are 36 OMs with 15 meters spacing attached to a vertical load-carrying cable called a string. The string is anchored to the bottom of the lake and streched by several buoys at the top. 12 OMs connected with 92 meters long cables to a control module form a section, i.e. a basic readout unit of the detector. The standard trigger condition requires a registration of two pulses on two neighboring OMs within the same section in a 100 ns time window with charges exceeding high ($\sim$ 4-5 p.e.) and low ($\sim$ 1.5-2 p.e.) channel dependent thresholds \cite{trigger2019}. There are 3 sections installed on each string. One central string is surrounded by 7 peripheral strings with a radius of 60\,m in a heptagonal formation. Alltogether, these 8 strings with 288 OMs form an independent unit -- cluster -- that is connected by a dedicated electro-optical cable to the shore. Once the trigger condition is fulfilled for any of the sections, a 5 $\mu$s event time frame is read out and stored from all the OMs in the cluster. The main source of events passing this trigger are atmospheric muons and random noise. The first cluster of the Baikal-GVD neutrino telescope was deployed in 2016 with 288 OMs. Until recent, there has been an annual increase of the deployed clusters resulting in 3\,456 OMs attached to 96 vertical strings (see Fig.(\ref{fig.GVDscheme})). The top and bottom OMs are placed at depths of 750\,m and 1\,275\,m, respectively.   

\section{Ambient light of Baikal water}

\begin{figure}[h]
\begin{center}$
\begin{array}{cc}
\includegraphics[width=67mm,height=40mm]{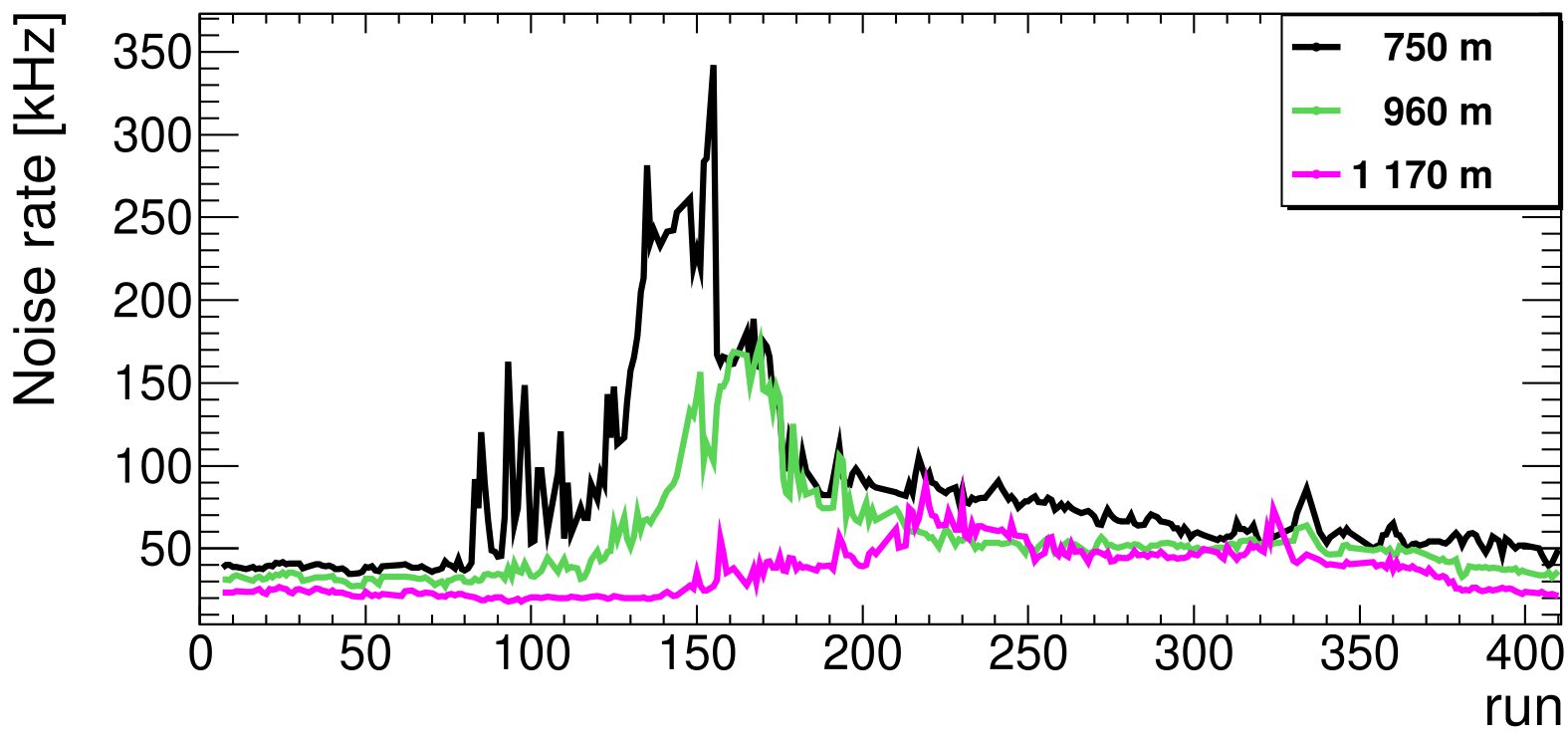}  &
\includegraphics[width=70mm,height=41mm]{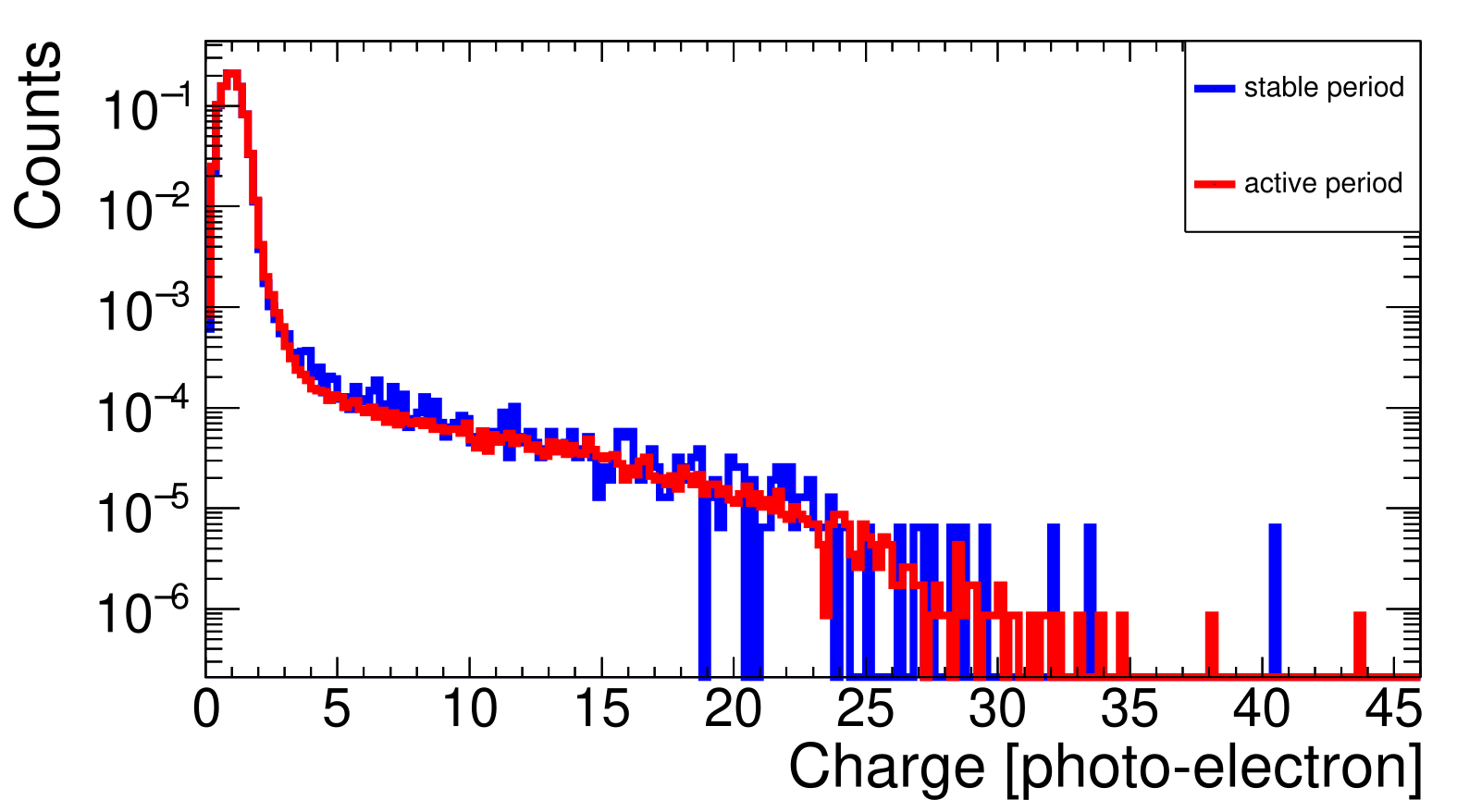}
\end{array}$
\end{center}
\caption{Left panel: Time dependence of PMT count rates on the same string at different depths. For the sake of simplicity, only 3 out of 36 OMs are shown, placed at depths of 750, 960, and 1\,170 meters. Right panel: Normalized charge distribution of detected pulses in units of photo-electrons for the same OM in an active (high-noise) period and in a stable (low-noise) period of optical activity. 
\label{fig.1}}
\end{figure}

One feature of the data acquistion system in the detector is that all pulses in a 5 $\mu$s time window from all 288 OMs of the cluster are stored as an event, once the trigger is fired by any section of the cluster. The times of triggering pulses lies in the middle of this window at $\sim$2.5 $\mu$s. The subject of our interest are pulses not related to the Cherenkov radiation, therefore, a cut on time of the pulses is used. It is set to select the first 2 $\mu$s of the 5 $\mu$s time frame, which is found to reduce the contribution of Cherenkov light to the overall pulse rate to a negligible level. The amount of the registered ambient light is proportional to the PMT rates inside each OM (the PMT dark current rate is neglected as it adds a minor contribution to the noise rate). As was shown in previous studies, the nature of these background photons is related to the luminescence of organic material suspended in the lake water \cite{nonstationarity,luminescence98}.  In Fig.\ref{fig.1} (left panel), time dependence of the PMT count rates recorded for selected OMs at three different depths for April 2022 -- April 2023 is shown. Here, the horizontal axis shows run numbers; a standard Baikal-GVD run lasts for $\sim$ 24 hours, representing a single uninterrupted mesurement series. Typical triggered event rate during the year is about 60-100 Hz. The data acquisition is practically continuous without substantial interputions during the year.
Two periods of relatively calm level of optical noise are intermitted by an increase of water luminescence. This increase occurs firstly on top OMs and advances slowly to bottom OMs as the luminescent layer moves downwards. The integrated charge distribution of pulses selected as the background noise is displayed in Fig.\ref{fig.1} (right panel). There is no significant difference in the shape of the distributions between the low-noise and high-noise periods of optical activity. Most of the pulses lie in the single p.e. peak. On average about 97$\%$ of pulses possess the charge lower than 2 p.e.. This behavior is crucial for higher level analysis of the Baikal-GVD data like track and cascade analyses, which are used for search of neutrino induced events. The contribution of the noise is naturally supressed by setting a lower limit on the charge of pulses registered by OMs of the detector. It is worth to note that the measurements are performed with a threshold of one third of a single photo-electron. The reason is substantial supression of the dark electronic noise of the PMTs. 
\begin{figure}[h]
\begin{center}$
\begin{array}{cc}
\includegraphics[width=65mm]{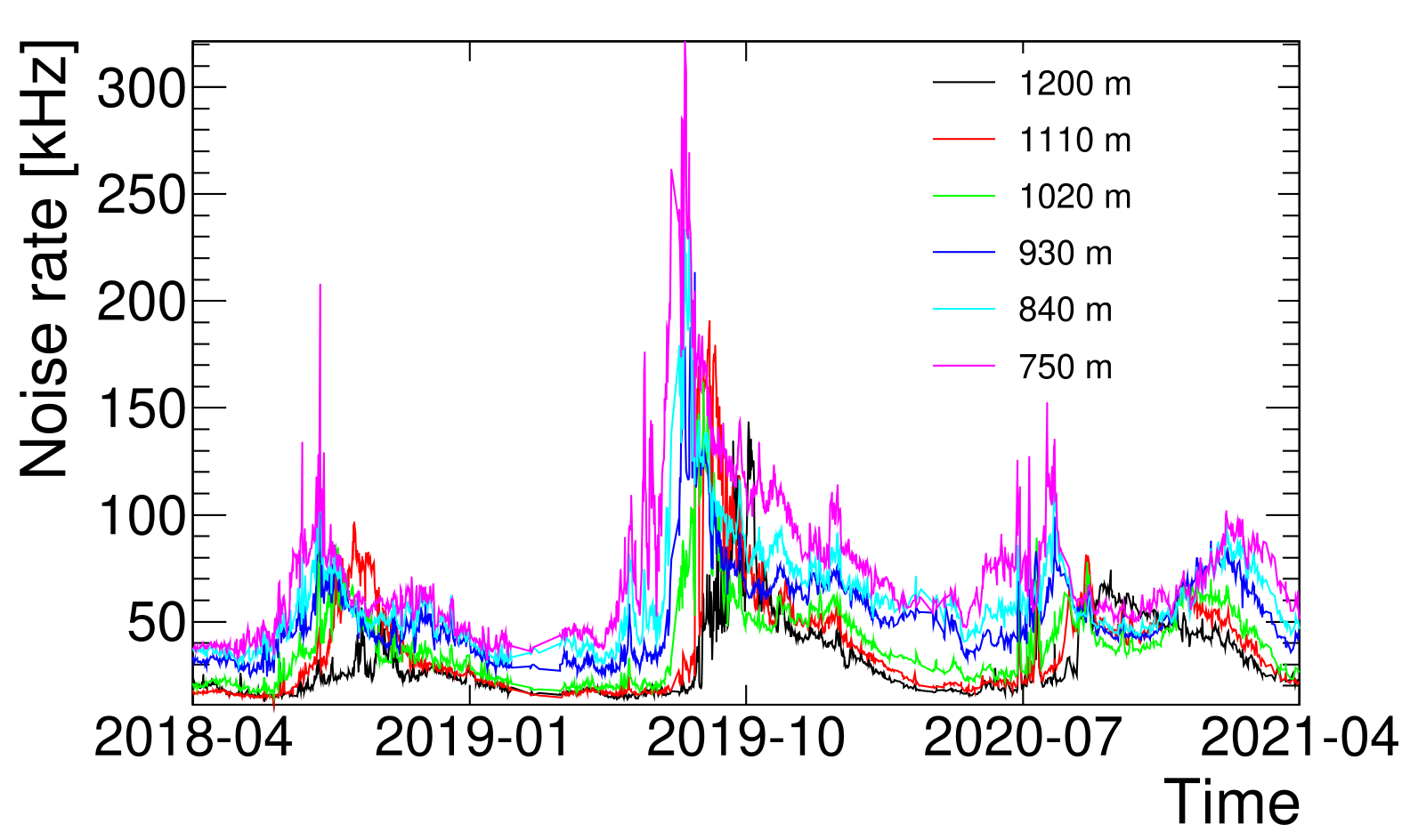}  &
\includegraphics[width=62mm]{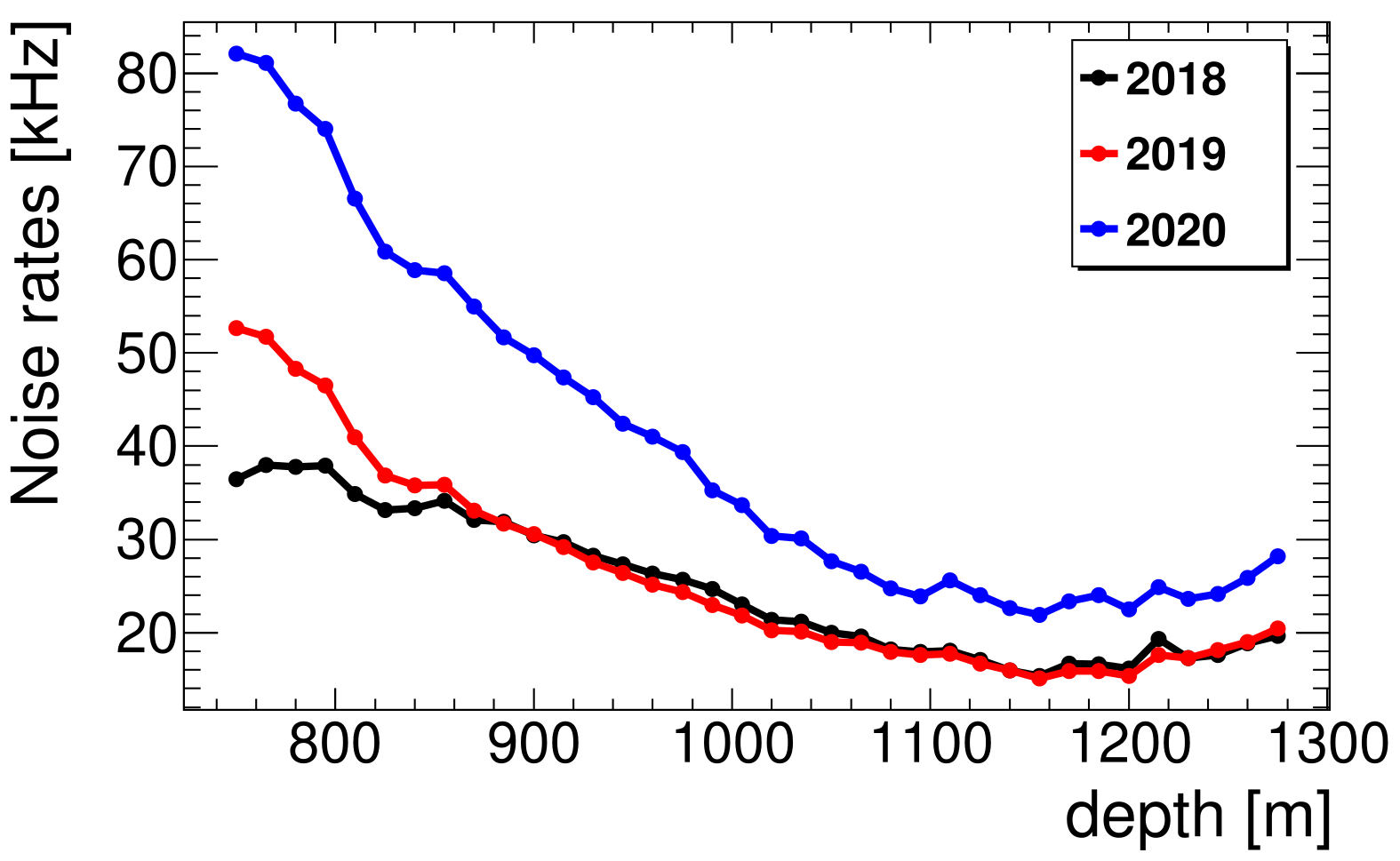}
\end{array}$
\end{center}
\caption{Left panel: PMT rates time dependence for OMs at the same string at different depths. Three years data are collected from April 2018 till April 2021. Right panel: Noise rates as a function of the depth. The PMT rates are averaged over 8 strings of a cluster. Data are from the stable periods from each of the years 2018, 2019, and 2020. The lake bed is at 1\,366 meters depth.
\label{fig.2}}
\end{figure}

\begin{figure}[h]
\begin{center}$
\begin{array}{cc}
\includegraphics[width=70mm]{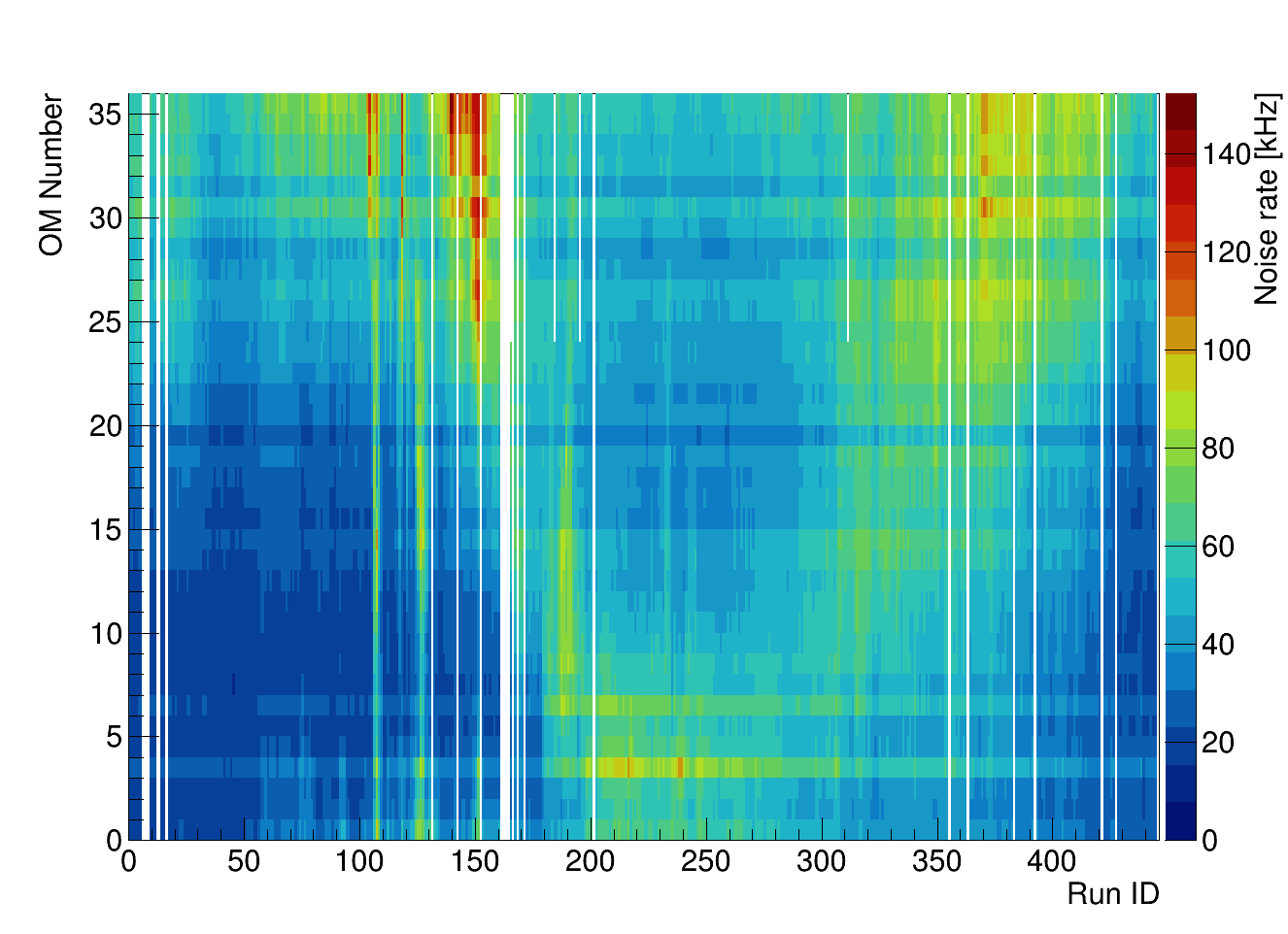}  &
\includegraphics[width=75mm]{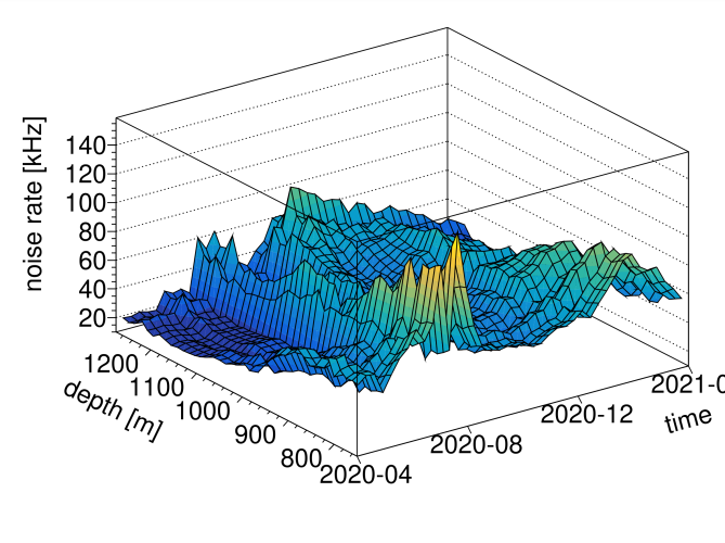}
\end{array}$
\end{center}
\caption{Left panel: Noise rates as a function of run number and OM position, shown for an example of string No.3 of cluster No.3 for the year 2020. The bottom OM number is labeled as 0 and the top one as 35. Right panel: The same as in the left panel only with OM number replaced with depth of the OM in meters and the run ID replaced with time.
\label{fig.3}}
\end{figure}

Time evolution of the noise rates for a period of three years for the cluster No.3 is displayed in Fig. \ref{fig.2} (left panel). Noise level in its maximum overshoots the calm level by a factor of 6 for the top OM and a factor of 2 for the bottom OM. By comparing the outbreak maximum at different depths, i.e. on different OMs on the same string, a vertical velocity profile of the luminescent layer is obtained (see \cite{nonstationarity}). In the beginning of August 2018, the estimated speed reached its maximal value of $\sim$16\,m/day, while it remained almost constant ($\sim$5\,m/day) till the end of September 2018, i.e. when the activity asymptotically reached the background plateau. There has been only one observation made of highly luminescent layer moving upwards in the year 2021 within the 3 years period of data taking (see Fig.\ref{fig.3}). The estimated speed reached its maximal value of $\sim$28\,m/day.

In general, the depth profile of the ambient light field is the same for all eight strings of a cluster. By taking the average of the noise rates for the OMs at the same horizon, 
the depth dependence of the background light noise is obtained. The average PMT rates versus the depth are presented in Fig.\ref{fig.2} (rigt panel). The input data are taken from the runs during the calm periods. The photon flux from the sunlight below a depth of $\sim$ 700 m is negligible as was shown in previous work \cite{luminescence98}. The turning point of the minimal PMT rates appears about 200 meters above the lakebed. A simple analytical approximation for the depth dependence of noise rates $f(H) \approx \exp(-H/H_0)$ is adopted from reference \cite{budnev2018}. Here, $f$ and $H$ are averaged noise rates in kHz and depth in meters, respectively. $H_0$ is the free parameter. A fitting procedure of the experimental data for the seasons 2018, 2019, and 2020, respectively, yield results of $H_0$ to be $461,~312$, and $292$ meters.

\begin{figure}[h]
\begin{center}$
\begin{array}{cc}
\includegraphics[width=68mm]{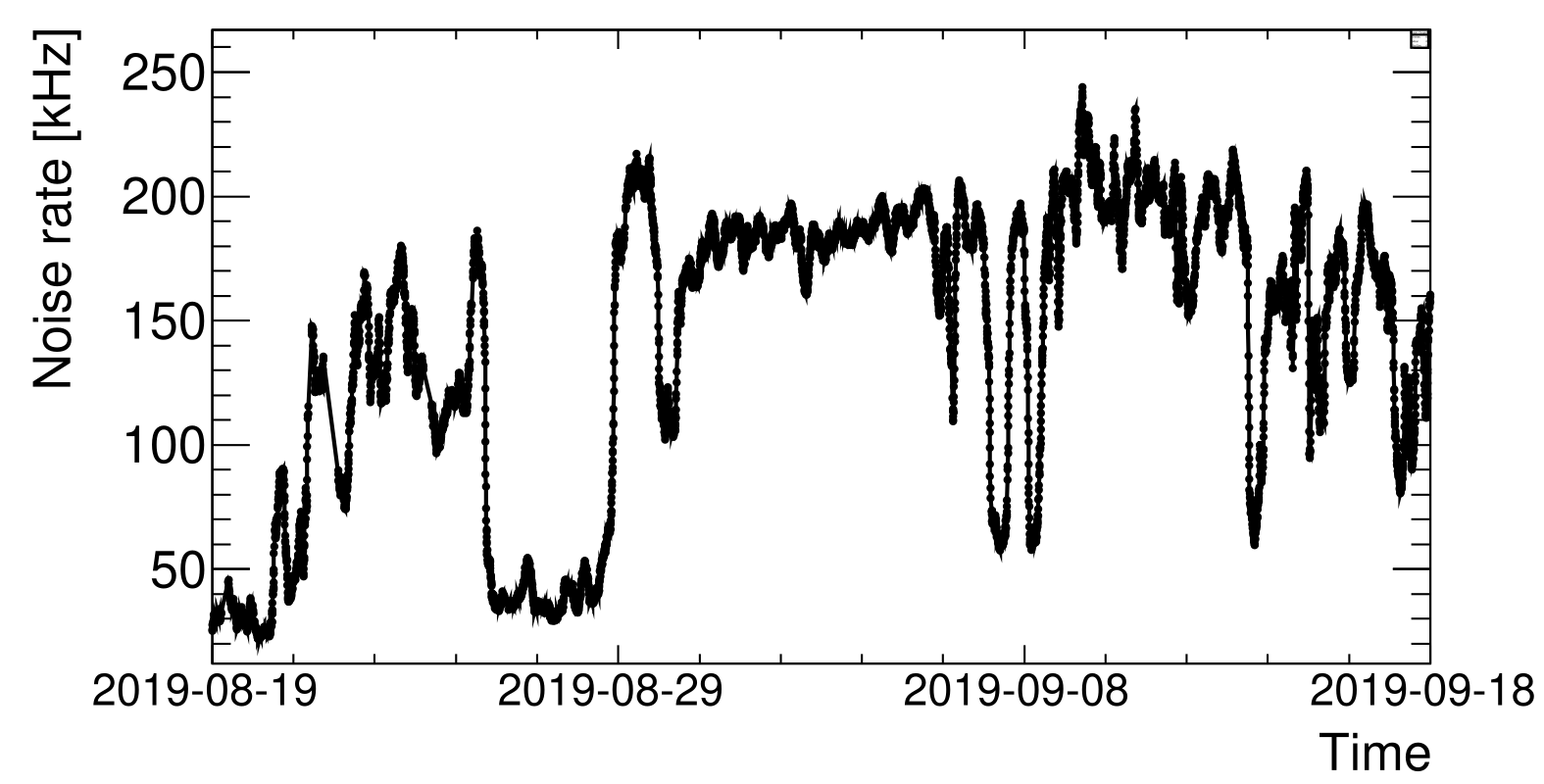}  &
\includegraphics[width=65mm,height=41mm]{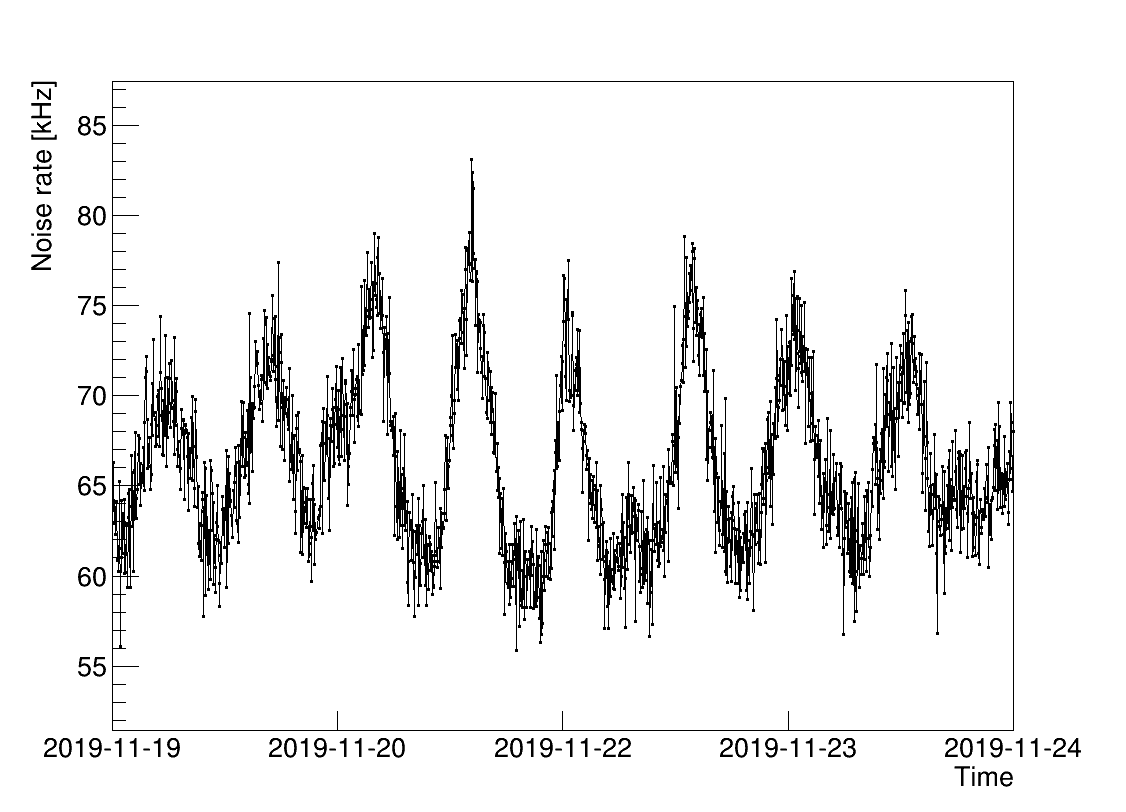}
\end{array}$
\end{center}
\caption{Left panel: Noise rates for a particular OM during the optically highly active period, namely from August till September 2019. A sudden outburst of the count rates is noticeable. Right panel: An example of the regular modulation of PMT rates. Data are from the period of stable plateau, namely from November 2019.   
\label{fig.4}}
\end{figure}

The time distribution of the PMT rates, as shown in Fig.\ref{fig.2} (left panel), exhibits sharp changes on top of relatively continuous smooth plateu. The effect is more visible in particularly selected time window displayed in Fig. \ref{fig.4} (left panel). The amplitude of these sudden changes reached almost 150\,kHz. The duration of such variations which distort the calm background lasts typically from several hours up to a few days. The effect is present in July -- September practically every year, i.e. the period of increased luminescent activity.This effect is in agreement with observations of our previous work \cite{LuminescenceICRC21}. The period of relatively stable plateau (for instance October -- February 2019) shows (see Fig. \ref{fig.4}, right panel) regular modulation of noise rates. The period of these modulations is quite stable with variations between 10 to 14 hours. 

\section{Conclusions}

Data on the luminescence of Baikal water collected with the Baikal-GVD neutrino telescope were presented. Annual increase of the luminescence activity between periods of relatively stable plateu has been reported. A single event of the luminescent layer propagating upwards has been observed in January 2021 with a maximum speed of 28\,m/day within the period 2018-2022 of data collection. The maximal amplitude of the sudden modulations in a highly active period reached the level of 150\,kHz.

\end{document}